# Generalized nanoscale electromagnetic boundary conditions and interfacial photonics

Yucheng Lai and Zhaona Wang[*]

Department of Physics, Applied Optics Beijing Area Major Laboratory, Beijing Normal University, Beijing, China, 100875

**Abstract**: Classical electromagnetic boundary conditions (EMBCs) fail to describe quantum interface phenomena at nanoscale. Here, we construct the interface model with a transition layer describing the electromagnetic field inhomogeneity across the interface. Generalized nanoscale EMBCs are derived by introducing the magnetic interfacial response functions (IRFs) $b_\perp, b_\parallel$ and are rewritten as three different forms based on Maxwell's equations in first order approximation. The corresponding Fresnel formula are further used to analyze the interfacial photonic phenomenon, demonstrating interesting behaviors of Brewster angle shifting, non-extinction at Brewster angle and distinctive non-classical absorption or gain effect at Brewster angle and the total internal reflection angles. IRFs-controlled GH-shift and angular GH-shift of Gaussian beam near Brewster angle are generated by the non-classical interface. These unique phenomena give us some guidance to measure the IRFs and expand interface photonics in nanoscale.

Electromagnetic boundary conditions (EMBCs) have very important applications in many branches of physics. They are the base of analyzing and determining the propagation behavior of electromagnetic (EM) waves and light waves on the interface, as well as the carrier transport characteristics on semiconductor interface. In traditional EM theory, the EMBCs are derived from the abrupt interface assumption and neglecting the integral contribution of EM field along the side wall of the integrating box[1, 2]. These EMBCs are generally used to describe the variation of the field intensity on both sides of the medium. But they break down and the related theoretical results demonstrate the gap with the experiments such as Nano-plasmonics[3-6], surface reflectance[7, 8] and surface photoexcitation[7] when we concern on the EM phenomena in the nanoscale. This gap might be induced by neglecting electron intrinsic scale, nonlocality property of the electric field response[9, 10] and the assumption of abrupt interface[7]. To narrow this gap, Feibelman[7] has developed the interfacial response function named as Feibelman *d* parameters as an effective approach for explaining the nanoscale EM phenomena[4, 7, 8]. In 2019, scientists further constructed the nanoscale EM framework based on the Feibelman *d* parameters[5], greatly promoting the development of the nanoscale electromagnetism. In 2020, Feibelman *d* parameters are extended to reveal the plasmon–emitter interactions at the nanoscale[11]. These works open the era of nanoscale electrodynamics for the metal-dielectric interface. However, the nanoscale EMBCs and interfacial response functions (IRFs) are given in a form with less physical fundamentals. The corresponding IRFs are difficult to be calculated and directly measured. There is an urgent need for constructing a simple physical model for its further development. Moreover, the reported EMBCs and the IRFs are for non-magnetic material under trivial EM field excitations. With the development of nano-photonics, magnetic response becomes more controllable through regulating the optical structures demonstrated as magnetic metamaterials [12, 13], negative permeability [14] and magnetic localized surface plasmons[15]. The generalized EMBCs including interfacial magnetic response are thus highly desired. In this letter, the generalized EMBCs are deduced based on the integral Maxwell's equations based on the interface response model with a transition layer. Two additional IRFs of $b_\perp, b_\parallel$ are introduced by considering the inhomogeneity of the magnetic fields within the transition layer. The obtained IRFs reflect the symmetry broken of the EM fields induced by the materials' response characteristics and distribution inhomogeneity across the interface, extending the connotation of the IRFs. Furtherly, the corresponding Fresnel formulas is proposed as a theoretical basis to analyze the reflection and refraction behaviors at the non-classical interface with transition layer. Unique non-extinction phenomenon at Brewster angle and Brewster angle shift are induced by the IRFs. The phase continuous variations and special GH-shifts are demonstrated across Brewster angle. Interestingly, the interface-induced absorption or gain effect is observed at Brewster angle and even at the total internal reflection angles. These unique properties supply the promising approach for directive measuring the IRFs at variety of interfaces, as well as open an era of regulating optical field at the nanoscale.

The interface is proposed at $z = 0$ plane and formed by the two isotropic bulk materials with permittivity and permeability of $\varepsilon_1$, $\mu_1$ $(z > 0)$ and $\varepsilon_2$, $\mu_2(z < 0)$, respectively. There is no free moving charge or current on the interface. The permittivity and permeability change continuously in a very thin transition layer from $z_{20}$ to $z_{10}$ as shown in Fig. 1a and Fig. S1c to demonstrate the special inhomogeneity of the EM field across the interface. The EM field in the non-transition region satisfying the traditional EMBCs of the sharp interface is looked as the zero-order approximate results. For an external stimulus of the harmonic planewave with angular frequency $\omega$, the integrating box in the nanoscale is constructed to obtain the EMBCs.

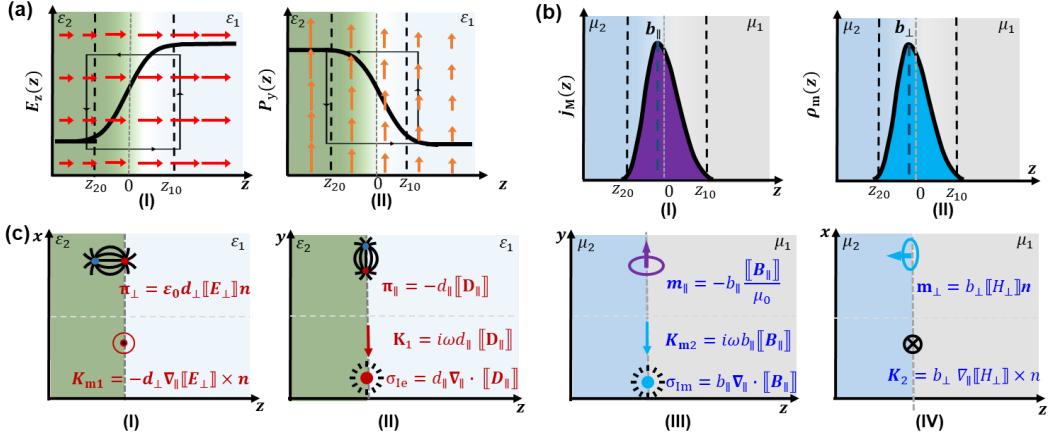

**FIG. 1. Physical models for nanoscale EMBCS.** (**a**) Continuous changes of electric field component $E_z$ (I) and polarization vector component $P_y$ (II) within the transition region. (**b**) The magnetic IRFs and the corresponding distributions of magnetization current density (I) and magnetization charge (II). (**c**) The equivalent non-classical interface model with interface polarization and magnetization dipole moments, the interface-induced polarization and magnetization charges , currents on the abrupt interface.

Using Maxwell's equations and considering the contributions of the transition layer as the first-order perturbation of the classical EMBCs from the abrupt interface, we obtain the generalized nanoscale EMBCs (Detailed information in section A in supporting Information (SI))

$$[\![E_\parallel]\!] = -d_\perp \nabla_\parallel [\![E_\perp]\!] - i\omega b_\parallel [\![B_\parallel]\!] \times n \tag{1a}$$

$$[\![D_\perp]\!] = d_\parallel \nabla_\parallel \cdot [\![D_\parallel]\!] \tag{1b}$$

$$[\![H_\parallel]\!] = -b_\perp \nabla_\parallel [\![H_\perp]\!] + i\omega d_\parallel [\![D_\parallel]\!] \times n \tag{1c}$$

$$[\![B_\perp]\!] = b_\parallel \nabla_\parallel \cdot [\![B_\parallel]\!] \quad . \tag{1d}$$

Here $[\![M_\alpha]\!] \equiv M_\alpha(z_{10}) - M_\alpha(z_{20})$ ($\alpha = \parallel, \perp$) represents the discontinuity of the tangential / normal component of field $M$ (an electric $E$, magnetic $B$, displacement $D$, or magnetizing $H$) across the interface. $\nabla_\parallel = \frac{\partial}{\partial x}e_x + \frac{\partial}{\partial x}e_y$ is the parallel component of Laplace operator. $n$ represents the unit vector normal to the interface from medium 2 to medium 1. $d_\perp, d_\parallel, b_\perp, b_\parallel$ are the defined IRFs for describing the discontinuity of the EM field at the nanoscale as

$$d_\perp \equiv \frac{\int_{z_{20}}^{z_{10}} z \frac{d\gamma}{dz} dz}{\varepsilon_B - 1} - \frac{z_1 \varepsilon_B - z_2}{\varepsilon_B - 1} \quad (2a)$$

$$d_\parallel \equiv \frac{\int_{z_{20}}^{z_{10}} z \frac{d\alpha}{dz} dz}{1 - \varepsilon_B} + \frac{z_2 \varepsilon_B - z_1}{1 - \varepsilon_B} \quad (2b)$$

$$b_\perp \equiv -\frac{z_1 \mu_B - z_2}{\mu_B - 1} + \frac{\int_{z_{20}}^{z_{10}} z \frac{d\delta}{dz} dz}{\mu_B - 1} \quad (2c)$$

$$b_\parallel \equiv \frac{z_2 \mu_B - z_1}{1 - \mu_B} + \frac{\int_{z_{20}}^{z_{10}} z \frac{d\beta}{dz} dz}{1 - \mu_B} \quad (2d)$$

where $\varepsilon_B = \varepsilon_2/\varepsilon_1$ and $\mu_B = \mu_2/\mu_1$ represent the relative permittivity and permeability of the two materials, respectively. Functions of $\gamma(z) = \frac{E_z(z)}{E_z(z_{20})}$, $\beta(z) = \frac{B_y(z)}{B_y(z_{10})}$, $\delta(z) = \frac{H_z(z)}{H_z(z_{20})}$ and $\alpha(z) = \frac{D_y(z)}{D_y(z_{10})}$ reflect the varying form of the EM fields within the transition region. The discontinuity of the EM field tangential component $\boldsymbol{E}_\parallel$ ($\boldsymbol{H}_\parallel$) across the interface is coupled not only with the normal component $E_\perp$ ($H_\perp$), but also with the inductive field component $\boldsymbol{B}_\parallel$ ($\boldsymbol{D}_\parallel$). The discontinuity of the inductive field ($D_\perp$, $B_\perp$) is proportional to the in-plane divergence of the corresponding tangential component. For the interface with uniform magnetic response, the generalized EMBCs are degenerated to the traditional nanoscale cases [5]. When the inhomogeneity of the EM fields across the interface can be neglected, the EMBCs are further degenerated to the traditional ones ($[\![\boldsymbol{E}_\parallel]\!] = 0$, $[\![\boldsymbol{D}_\perp]\!] = 0$, $[\![\boldsymbol{H}_\parallel]\!] = 0$, $[\![\boldsymbol{B}_\perp]\!] = 0$).

Formulas (2) give us an opportunity to calculate the IRFs for the special interface with different variation forms of α, γ, β, δ, which might be originated from the EM response inhomogeneity, local adsorption charges, special external field-generated polarization or magnetizations. Considering the localization characteristics of the interface, we extend the integral region by performing coordinate system translation transformation (Detailed information in SI B1-B5) and rewrite the IRFs as

$$d_\perp = \frac{\int_{-\infty}^{+\infty} z \frac{dE_z}{dz} dz}{[\![E_z]\!]} = \frac{\int_{-\infty}^{+\infty} z \rho_{\text{ind}} dz}{\int_{-\infty}^{+\infty} \rho_{\text{ind}} dz} \quad (3a)$$

$$d_\parallel = \frac{\int_{-\infty}^{+\infty} z \frac{dD_y}{dz} dz}{[\![D_y]\!]} = \frac{\int_{-\infty}^{+\infty} z \frac{d}{dz} j_{\text{py}} dz}{\int_{-\infty}^{+\infty} \frac{dj_{\text{py}}}{dz} dz} \quad (3b)$$

$$b_\perp = \frac{\int_{-\infty}^{+\infty} z \frac{dH_z}{dz} dz}{[\![H_z]\!]} = \frac{\int_{-\infty}^{+\infty} z \rho_m dz}{\int_{-\infty}^{+\infty} \rho_m dz} \quad (3c)$$

$$b_\parallel = \frac{\int_{-\infty}^{+\infty} z \frac{dB_y}{dz} dz}{[\![B_y]\!]} = \frac{\int_{-\infty}^{+\infty} z j_M dz}{\int_{-\infty}^{+\infty} j_M dz} \tag{3d}$$

Among them, $\rho_{\text{ind}}$ and $\rho_{\text{m}}$ represent the field-gradient-induced local polarization charge density and equivalent magnetization charge density within the transition layer, and $j_M/j_{\text{py}}$ are the local magnetization/polarization current density across the interface. $d_\perp$ and $b_\perp$ denote the centroids of interface-induced polarization charge and magnetization charge, respectively. $d_\parallel$ and $b_\parallel$ denote the centroid of normal derivative of tangential polarization current density which results from time-dependent displacement field and the centroid of magnetization current density (Fig. 1b and Fig. S2d), respectively.

It is obvious that the IRFs come into being due to symmetry breaking on normal/tangential EM fields across the interface. On an ideal charge-neutral infinite plane, $d_\parallel \to 0$ is related to the in-plane homogeneous polarization field at the traditional interface [14]. But $d_\parallel$ should be considered when the in-plane polarization field is inhomogeneous from local self-polarization and strain-induced polarization[16, 17]. For the interface formed by magnetic materials, $b_\parallel$ and $b_\perp$ induced by the magnetic response inhomogeneous cannot be neglected. Moreover, nontrivial stimuli EM field supplies an alternative approach to excite local interfacial magnetization through coupling angular momentum of EM waves to atoms and molecules in the transition region.

According to the definition of electric dipole moment and magnetic dipole moment, the proposed non-classical interface-induced electric dipole moment ($\boldsymbol{\pi}_\perp$ and $\boldsymbol{\pi}_\parallel$) and magnetic dipole moment ($\boldsymbol{m}_\perp$ and $\boldsymbol{m}_\parallel$) are further introduced to describe the discontinuity of the EM fields across the interface (Detailed information in Section B in SI). And these dipole moments are defined as

$$\boldsymbol{\pi}_\perp = \int_{-\infty}^{+\infty} z \rho_{\text{ind}} dz\, \boldsymbol{n} = \varepsilon_0 d_\perp [\![E_\perp]\!] \boldsymbol{n} \tag{4a}$$

$$\boldsymbol{\pi}_\parallel = -d_\parallel [\![\boldsymbol{D}_\parallel]\!] \tag{4b}$$

$$\boldsymbol{m}_\perp = \int_{-\infty}^{+\infty} z \rho_{\text{m}} dz\, \boldsymbol{n} = b_\perp [\![H_\perp]\!] \boldsymbol{n} \tag{4c}$$

$$\boldsymbol{m}_\parallel = \int_{-\infty}^{+\infty} z \boldsymbol{j}_M dz = -b_\parallel \frac{[\![\boldsymbol{B}_\parallel]\!]}{\mu_0} \quad . \tag{4d}$$

The introduced $\boldsymbol{\pi}_\parallel$ is originated from the normal discontinuity of in-plane electric dipole momentum, vanishing in neutral strictly planar interfaces[8, 10]. But it should be considered for surface roughness[18], bound screening[9] and inhomogeneous in-plane polarization. This is beyond the meaning of the interface model presented in previous articles [4, 5]. The proposed magnetization charge and interface magnetization current within the transition layer can be equivalent as the corresponding interface magnetic dipole moment $m_⊥/\boldsymbol{m}_∥$ at the interface. Therefore, the generalized nanoscale EMBCs can be expressed as

$$[\![E_∥]\!] = -(\frac{-1}{\varepsilon_0}\nabla \times \boldsymbol{\pi}_⊥ + \mu_0 \frac{\partial \boldsymbol{m}_∥}{\partial t}) \times \boldsymbol{n} \tag{5a}$$

$$[\![H_∥]\!] = (\frac{\partial \boldsymbol{\pi}_∥}{\partial t} + \nabla \times \boldsymbol{m}_⊥) \times \boldsymbol{n} \tag{5b}$$

$$[\![D_⊥]\!] = -\nabla_∥ \cdot \boldsymbol{\pi}_∥ \tag{5c}$$

$$[\![B_⊥]\!] = -\mu_0 \nabla_∥ \cdot \boldsymbol{m}_∥ \tag{5d}$$

The discontinuity of in-plane electric (magnetic) field is relevant to curl of normal component of non-classical interface electric (magnetic) dipole moment and time-variation rate of the tangential components of the interface magnetic (electric) dipole moment. The discontinuity of vertical components of electric displacement vector and magnetic induction intensity are dependent on the in-plane divergence of the tangential components of the interface electric and magnetic dipole moment, respectively. It tells us that the interface polarization and magnetization processes are the source of the EM field discontinuity. And the non-classical interface with a transition layer can be simply looked as the abrupt interface with interface-induced electric and magnetic dipole moments.

Based on the physical meaning of interface dipole moment, the linear density of interface polarization current $\boldsymbol{K_1}$ and magnetization current $\boldsymbol{K_2}$ is defined as variation rate of in-plane dipole moment $\boldsymbol{\pi}_∥$ with time and curl of the interface-induced magnetic dipole moment $\boldsymbol{m}_⊥$, respectively. The density of interface equivalent magnetic charge currents $\boldsymbol{K_{m1}}$ and $\boldsymbol{K_{m2}}$ is related to curl of the interface-induced dipole moment $\boldsymbol{\pi}_⊥$ and time-varying rate of in-plane magnetic dipole moment $\boldsymbol{m}_∥$, respectively. Interface equivalent polarization charge sheet density $\sigma_{Ie}$ and equivalent magnetization charge sheet density $\sigma_{Im}$ originates from divergency of the in-plane electric and magnetic dipole momentum $\boldsymbol{\pi}_∥$ and $\boldsymbol{m}_∥$, respectively. Detailed information is in section C in SI. Their definition formula are as follows (Fig. 1c)

$$\boldsymbol{K_1} = \frac{\partial \boldsymbol{\pi}_∥}{\partial t} = i\omega d_∥ [\![\boldsymbol{D}_∥]\!] \tag{6a}$$

$$\boldsymbol{K_2} = \nabla \times \boldsymbol{m}_⊥ = b_⊥ \nabla_∥[\![H_⊥]\!] \times \boldsymbol{n} = \nabla_∥ m_⊥ \times \boldsymbol{n} \tag{6b}$$

$$\boldsymbol{K}_{\mathbf{m1}} = -\frac{1}{\varepsilon_0} \nabla \times \boldsymbol{\pi}_\perp = -d_\perp \nabla_\| [\![E_\perp]\!] \times \boldsymbol{n} \tag{6c}$$

$$\boldsymbol{K}_{\mathbf{m2}} = \mu_0 \frac{\partial \boldsymbol{m}_\|}{\partial t} = i\omega b_\| [\![\boldsymbol{B}_\|]\!] \tag{6d}$$

$$\sigma_{\text{Ie}} = \frac{1}{i\omega} \nabla_\| \cdot \boldsymbol{K}_\mathbf{1} = -\nabla_\| \cdot \boldsymbol{\pi}_\| = d_\| \nabla_\| \cdot [\![\boldsymbol{D}_\|]\!] \tag{6e}$$

$$\sigma_{\text{Im}} = -\mu_0 \nabla_\| \cdot \boldsymbol{m}_\| = \frac{1}{i\omega} \nabla_\| \cdot \boldsymbol{K}_{\mathbf{m2}} = b_\| \nabla_\| \cdot [\![\boldsymbol{B}_\|]\!]. \tag{6f}$$

Therefore, the nanoscale EMBCs can also be expressed as the equivalent polarization and magnetization charge sheet density on the interface, the interface-induced current and magnetic charge current as

$$[\![\boldsymbol{E}_\|]\!] = -(\boldsymbol{K}_{\mathbf{m1}} + \boldsymbol{K}_{\mathbf{m2}}) \times \boldsymbol{n} \tag{7a}$$

$$[\![\boldsymbol{H}_\|]\!] = (\boldsymbol{K}_\mathbf{1} + \boldsymbol{K}_\mathbf{2}) \times \boldsymbol{n} \tag{7b}$$

$$[\![D_\perp]\!] = \sigma_{\text{Ie}} \tag{7c}$$

$$[\![B_\perp]\!] = \sigma_{\text{Im}}. \tag{7d}$$

Compared to the traditional EMBCs of the abrupt interface with the free charge sheet density $\sigma_e$ and the free current line density $\boldsymbol{\alpha}_f$ ($[\![\boldsymbol{E}_\|]\!] = 0$, $[\![\boldsymbol{H}_\|]\!] = \boldsymbol{\alpha}_f \times \boldsymbol{n}$, $[\![D_\perp]\!] = \sigma_e$, $[\![B_\perp]\!] = 0$), the role of interface-induced polarization current $\boldsymbol{K}_\mathbf{1}$ and magnetization current $\boldsymbol{K}_\mathbf{2}$ is equivalent to the free current line density $\boldsymbol{\alpha}_f$. The role of interface-induced charge sheet density $\sigma_{\text{Ie}}$ is equivalent to the $\sigma_e$. But the interface-induced magnetic charge current $\boldsymbol{K}_{\mathbf{m1}}$ and $\boldsymbol{K}_{\mathbf{m2}}$ and magnetization charge sheet density $\sigma_{\text{Im}}$ are introduced. Equations (7) are thus both symmetric and identical to the EM field parameters when considering the interface contributions as polarization and magnetization charges and currents on the interface, demonstrating the high duality of electrical and magnetic properties. It also indicates that contributions of the interface EM field discontinuity can be equivalent as the amount polarization and magnetization charges, current and magnetic charge current on the abrupt interface. This is the fundamental of constructing the equivalent interface model of abrupt interface with the polarization and magnetization charges and currents for calculations.

We consider the harmonic plane wave with angular frequency $\omega$ and vacuum wave number $k_0$ incident on the interface with the relative permittivity discontinuity $\Delta\varepsilon = \varepsilon_{r2} - \varepsilon_{r1}$ and relative permeability discontinuity $\Delta\mu = \mu_{r2} - \mu_{r1}$ at the incident angle of $\theta_1$ and refraction angle of $\theta_2$. Based on the generalized nanoscale EMBCs, the retarded Fresnel formula are obtained as

(Detailed information in Section D in SI)

$$r_p = \frac{(Z_1 \cos\theta_1 - Z_2 \cos\theta_2) + i\Delta\varepsilon Z_1 Z_2 k_0 (Ad_\perp - Bd_\parallel) - ik_0 b_\parallel \Delta\mu}{(Z_1 \cos\theta_1 + Z_2 \cos\theta_2) - i\Delta\varepsilon Z_1 Z_2 k_0 (Ad_\perp + Bd_\parallel) + ik_0 b_\parallel \Delta\mu} \quad (8a)$$

$$t_p = \frac{2Z_1 \cos\theta_1}{(Z_1 \cos\theta_1 + Z_2 \cos\theta_2) - i\Delta\varepsilon Z_1 Z_2 k_0 (Ad_\perp + Bd_\parallel) + ik_0 b_\parallel \Delta\mu} \quad (8b)$$

$$r_s = \frac{\left(Z_2 \frac{1}{\cos\theta_2} - Z_1 \frac{1}{\cos\theta_1}\right) + i\Delta\mu k_0 (Ab_\perp/B - b_\parallel) - ik_0 d_\parallel \frac{Z_1 Z_2}{B} \Delta\varepsilon}{\left(Z_2 \frac{1}{\cos\theta_2} + Z_1 \frac{1}{\cos\theta_1}\right) - i\Delta\mu k_0 (Ab_\perp/B + b_\parallel) + ik_0 d_\parallel \frac{Z_1 Z_2}{B} \Delta\varepsilon} \quad (8c)$$

$$t_s = \frac{2Z_2 \frac{1}{\cos\theta_2}}{\left(Z_2 \frac{1}{\cos\theta_2} + Z_1 \frac{1}{\cos\theta_1}\right) - i\Delta\mu k_0 (Ab_\perp/B + b_\parallel) + ik_0 d_\parallel \frac{Z_1 Z_2}{B} \Delta\varepsilon} \quad (8d)$$

Among those, we define dimensionless wave impedance as $Z_i = \sqrt{\mu_{ri}/\varepsilon_{ri}}$ (i=1,2), parameters $A = \sin\theta_1 \sin\theta_2$ and $B = \cos\theta_1 \cos\theta_2$, amplitude reflection coefficient $r_s$ and $r_p$, amplitude transmission coefficient $t_s$ and $t_p$. The Fresnel formula based on the nanoscale EMBCs is related to the IRFs. For p-waves, the reflection and transmission coefficients of $r_p$ and $t_p$ are both related to the IRFs of $d_\perp, d_\parallel$ and $b_\parallel$, while the coefficients $r_s$ and $t_s$ of the s-waves are related to the IRFs of $b_\perp, b_\parallel$ and $d_\parallel$. Meanwhile, the contribution of the IRFs is modified by the discontinuity of permittivity $\Delta\varepsilon$ (permeability $\Delta\mu$) across the interface, indicating the field-discontinuity origin of the IRFs. This dependence relation endows a potential way to separately modify the s-waves and p-waves by tuning the corresponding IRFs. When the IRFs are zero, the Fresnel formula degenerated to the traditional ones in flat surface in most cases[19].

The relative contributions of the IRFs on the Fresnel coefficients is related to the incident angle. And the IRFs will contributes a lot to the amplitude reflective coefficient around Brewster angle ($\theta_B$) and total internal reflection angles. The corresponding ratio of $r_p$ ($r_s$) to classical one $r_{pc}$ ($r_{sc}$) at $\theta_B$ approaches maximum so the contribution of non-classical amendment is dominated (Detailed information in section E in SI). To display the influence of $d_\perp/d_\parallel$ on $r_p$, we choose the vacuum-SiO$_2$ interface formed by non-absorptive materials with $\varepsilon_{r2} = 2.25, \mu_{r2} = 1$. The variations of the modulus ($|r_p|$) and argument ($\varphi_p$) of $r_p$ with the incident angle are shown in Fig.2 and Fig. S2 under different $d_\perp/d_\parallel$. A non-extinction phenomenon at $\theta_B$ is resulted from the real part of $d_\perp$ (Re($d_\perp$)) on the interface formed by non-absorptive materials (Fig. 2a). This is different from the result of the traditional Fresnel formulas based on the abrupt interface, but agree with the experimental results[7]. The position shifting $\Delta\theta_{B1}$ of Brewster angle is caused by the imagery part of $d_\perp$ (Im($d_\perp$)). And $\Delta\theta_{B1}$ is proportional to $k_0(-\text{Im}(d_\perp) + \text{Im}(d_\parallel))$ and becomes larger as increasing $\varepsilon_{r2}$ (Fig. S3a). A typical $\Delta\theta_{B1}$ can reach 0.1°~1.0° in an observable range. Besides, the coefficient $|r_{p,min}|$ at $\theta_B$ is proportional to $k_0|\text{Re}(d_\perp) - \text{Re}(d_\parallel)|$ and large $|\Delta\varepsilon|$ can enhance such non-classical phenomenon of non-extinction (Section E and Fig. S3 in SI). This non-classical $|r_{p,min}|$ is typically $10^{-6}$~$10^{-5}$. The reflection intensity is easily detected based on the current technique for a watt or milliwatt laser stimuli. The phase variation $\varphi_p$ of the reflected light relative to the incident light changes smoothly with incident angle around $\theta_B$, different from the phase jump at $\theta_B$ predicted by classical EMBCs (Fig. 2b). The slope of $\varphi_p - \theta$ curve is proportion to $1/\text{Re}(d_\perp)$ at $\theta_B$ where $\varphi_p = \pm\frac{\pi}{2}$. The sign of $\text{Re}(d_\perp)$ determines the phase variation range and sign of $\text{Im}(d_\perp)$ affect the shape of phase variations. As a result, phase variation $\varphi_p$ could be modified in $[0,2\pi]$ by smartly controlling the incident angle around $\theta_B$ and tuning the IRF $d_\perp$ at the nanoscale. The IRFs $b_\perp$ and $b_\parallel$ demonstrate similar tuning role on the reflection coefficient $r_s$ (Fig. S4). This effect supplies a possible method to separately tuning the wave front of the s-waves and p-waves by controlling the corresponding IRFs on the interface.

When the incidental angle is larger than the critical angle, the IRF-induced non-classical absorption or gain effect will be observed as $|r_p|$ smaller or larger than unit on the interface formed by non-absorption materials (Fig. 2c). This unique phenomenon is totally originated from the non-classical interface properties at the nanoscale. $\text{Im}(d_\perp) > 0$ or $\text{Im}(d_\parallel) < 0$ means interface-induced absorption effect. Interestingly, $\text{Im}(d_\perp) < 0$ or $\text{Im}(d_\parallel) > 0$ will introduce unique gain effect on the traditional interface. This tuning role of non-classical absorption or gain effect from the interface can also be observed on the noble metal Ag ($\varepsilon_{r2} = -198.189 + 6.76i, \mu_{r2} = 1$) surface[20] with strong absorption (Fig. 2d and Fig. S5). This interface effect might give us a chance to achieve PT symmetry[21] photonics at the nanoscale by controlling the imagery part of IRFs. Notably, the non-classical effects of $d_\perp$ are largest (blue) while those of $d_\parallel$ are weakest (black) at critical angle, supplying a new way to distinguish and measure $d_\perp$ and $d_\parallel$ on the interface. Particularly, this non-classical gain effect might generate amazing extinction phenomenon for the weak-absorption-material-formed interface with a proper IRF by strong coupling the bulk absorption and the interface gain effect generated by the interface-induced dipole moment (Fig. S6). And a phase jump is observed as the absence of transition layer.

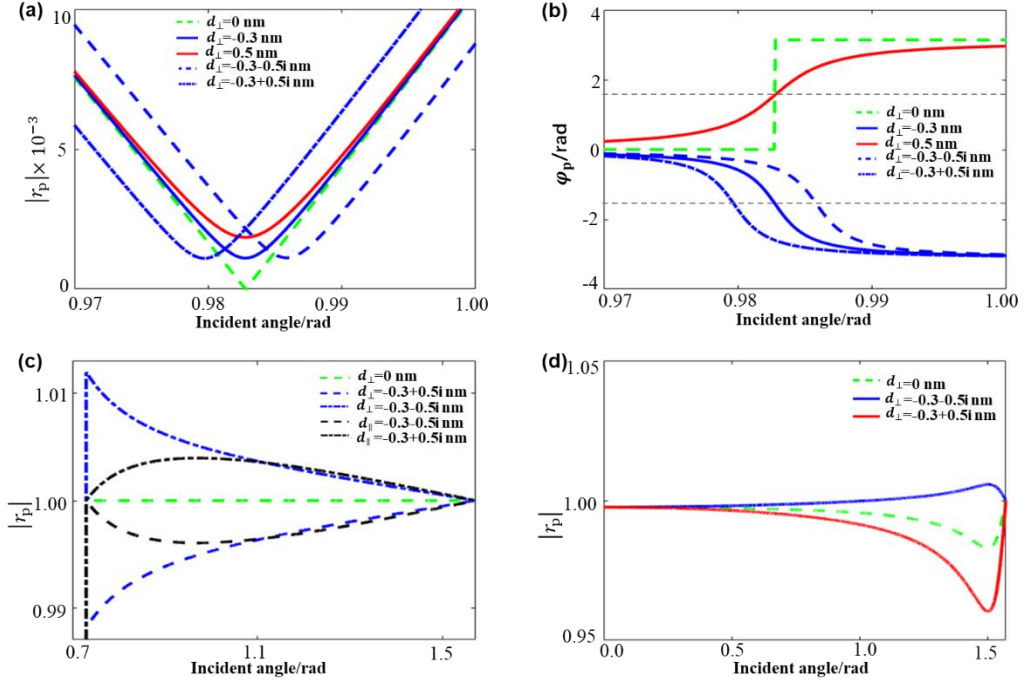

**FIG. 2.** Reflection coefficient $r_p$ variations with incident angle at the interface with different IRFs $d_\perp/d_\parallel$. (a, b) $|r_p|$ (a) and $\varphi_p$ (b) of the vacuum-SiO$_2$ interfaces with different $d_\perp$ vary with the incident angle near $\theta_B$ for the incident plane wave with $\lambda$=589 nm. (c) $|r_p|$ on the SiO$_2$-vacuum interfaces with different $d_\perp/d_\parallel$ changes with the incident angle larger than the critical angle. (d) $|r_p|$ on the vacuum-Ag interfaces with different $d_\perp$ varies with incident angle for the plane wave with $\omega$=0.64 eV.

The continuous and fast phase shift around Brewster angle means a large Goos-Hänchen (GH) shift. We suppose a Gaussian beam incident on the dielectric interface at $z = 0$ with an incident angle $\theta_i$ in Fig. 3a. The GH-shift around $\theta_B$ is calculated by $\Delta_{GH} = -\frac{1}{k}\frac{d\varphi_p}{d\theta_i}$ which is provided by Artmann [22]. A large positive/negative GH shift is obtained on the non-absorption material-formed interface with different $d_\perp$ (Fig. 3b) and $d_\parallel$ (Fig. S7a, SI). This GH shift is totally induced by the interface-generated non-classical dipole on the interface, different from the bulk weak absorption-induced GH-shift[23]. Interestingly, the sign of $\Delta_{GH}$ can be flexibly controlled by the sign of Re($d_\perp$) and Re($d_\parallel$). This phenomenon indicates the rich GH-shift behaviors including large positive or negative GH-shift on the non-absorptive interface with different IRFs. Moreover, $\Delta_{GH}$ is inversely proportional to $k_0(\text{Re}(d_\perp) - \text{Re}(d_\parallel))$ (Detailed information in Section F in SI). We can observe large GH-shift under the conditions of small $\varepsilon_{r2}$ and large $\varepsilon_{r1}$ at a fixing Re($d_\perp$) − Re($d_\parallel$) (Fig. S7b), indicating a direction for large GH-shift at Brewster angle. This large GH-shift caused by the EM field variation within the transition layer can also be found at the non-classical interface with magnetic metamaterials (Fig. S7c, d) and weak-absorption material (Fig. S8).

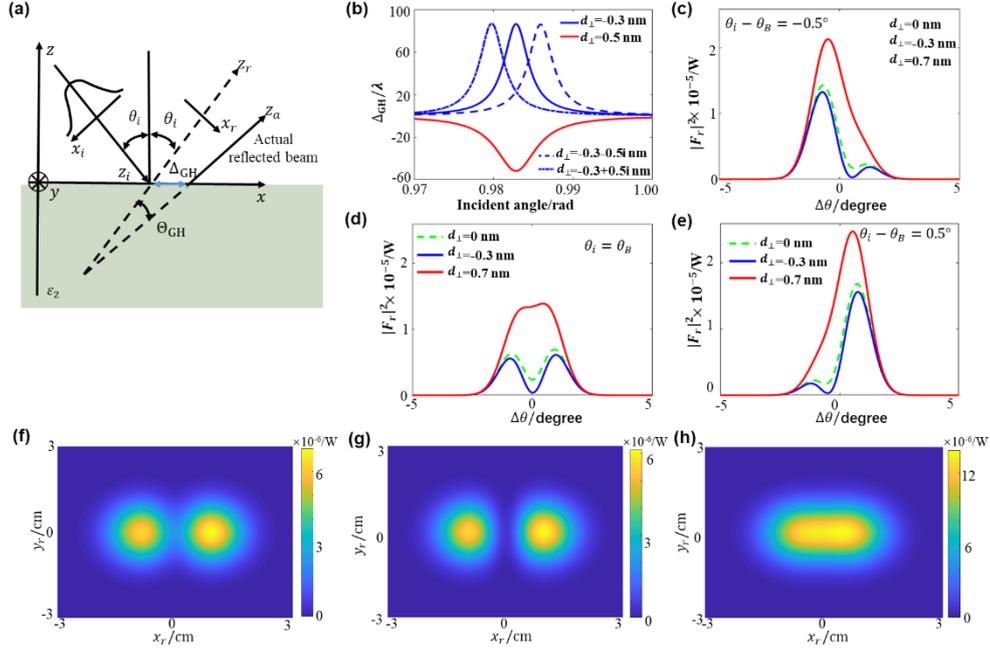

**FIG. 3. Effect of $d_\perp$ on GH-shift and angular distribution**. (**a**) Local Cartesian coordinate systems of $xyz$ for the interface, $x_i y z_i$ for incident beams and $x_r y z_r$ ($x_a y z_a$) for the virtual (actual) reflected beams. The direction $\hat{z}_r$ indicates the geometrical ray axis of the reflected beam based on the Snell formula and $\hat{z}_a$ represents the actual beam reflected beam. The angle between $\hat{z}_r$ and $\hat{z}_a$ are GH angular shift $\Theta_{GH}$. (**b**) GH-shift varies with the incident angle $\theta_i$ near $\theta_B$ when the Gaussian beam with $\lambda$ =589 nm incidents on the vacuum-SiO$_2$ interface with different $d_\perp$. (**c-e**) Angular distribution of the reflected Gaussian beam in the $x_r y$ plane on the vacuum-Si interface with different $d_\perp$ at different incident angles of $\theta_i = \theta_B - 0.5°$ (**c**), $\theta_i = \theta_B$ (**d**) and $\theta_i = \theta_B + 0.5°$ (**e**). (**f-h**) Intensity distribution of the reflected beam on the plane of $z_r = 50$ cm under $d_\perp = 0$ (**f**), $d_\perp = -0.3$ nm (**g**) and $d_\perp = 0.7$ nm (**h**). The incident Gaussian beam has a waist of $w_0 = 6$ μm and the wavelength of $\lambda$=449.2 nm.

When linearly polarized Gaussian beam incidents on the weak absorption materials, angular GH-shift $\Theta_{GH}$ and angular distribution can be obtained by the formulas provided by Chan[24]. Angular distribution of reflected Gaussian beam near $\theta_B$ is demonstrated in Fig. 3c-e. Classically, the reflected beam will be divided into two symmetric wave packets at two sides of $\theta_B$ when the incident angle is $\theta_i=\theta_B$. When the beams incident at $\theta_i>\theta_B$ or $\theta_i<\theta_B$, two asymmetric wave packets are observed, following a large angular GH-shift near $\theta_B$ (Fig. 3c, 3e, green lines). The intensity ratio of the two wave packets could be tuned by the IRF $d_\perp$. Interestingly, amalgamation phenomenon of the two wave packets can be induced by the non-classical interface dipole (Fig. 3c-e, red lines). This unique behavior might be attributed to radiation of surface dipole. This phenomenon can also be be induced by the IRF $d_\parallel$ (Fig. S9a-c) on the weak-absorption interface. For the non-absorptive interfaces, a relative weak non-classical effect is observed (Fig. S9d-f). The corresponding intensity distributions of the reflected beams on the plane $z_r$=50 cm are shown in Fig. 3f-h. Intensity distribution changes enormously with the real part of $d_\perp$. This phenomenon offers us a chance to measure $d_\perp$ at the plane by detecting angular distribution of the reflected light.

In this letter, a non-classical interface model with EM field transition layer is constructed through a classical view for describing the interfacial non-classical physical processes. The generalized nanoscale EMBCs are developed based on the integral Maxwell equations. On this basis, four IRFs $d_\perp, d_\parallel$, $b_\perp, b_\parallel$ are presented with clear physical meanings beyond the Feibelman parameters. To reveal the physical pictures of non-classical interface response, the model of interfacial dipole moment and non-classical interface-induced polarization and magnetization charge, current and magnetic charge current are further built up. And the generalized EMBCs are rewritten in two more symmetric forms, demonstrating the duality of electricity and magnetism. The generalized EMBCs provide a necessary theoretical basis for researching the EM processes at the nanoscale and further revealing the optical properties of nanoscale interface with magnetic response characteristics. As an application of these theory, we develop the retarded Fresnel formulas and investigate non-classical phenomena induced by the IRFs. The unique non-extinction phenomenon at $\theta_B$ and Brewster angle shifting are induced by the non-classical interface with the non-absorption materials. The magical absorption effect and gain effect can be observed around the critical angle. And interesting merging behavior of two wave packets from reflected Gaussian beam is induced by the IRF. These unique properties demonstrate the rich physical processes generated by the non-classical interface, as well as provide some feasible approaches to measure and distinguish the IRFs on the interfaces. Our results supply a basic tool to reveal the EM phenomena at the nanoscale and might promote the interface photonics in the nanoscale in near future.